\documentclass{PoS}

\usepackage{listings}
\usepackage{multicol}
\usepackage{epstopdf}
\usepackage{amsfonts,amsmath,amssymb}
\usepackage{mathtools}
\usepackage[font=small,labelfont=bf]{caption}

\title{DDalphaAMG for Twisted Mass Fermions}

\ShortTitle{DD-$\alpha$AMG for TM fermions}

\author{\speaker{Simone Bacchio}\textsuperscript{, a, b}, Constantia Alexandrou\textsuperscript{a,c},
Jacob Finkenrath\textsuperscript{c},\newline Andreas Frommer\textsuperscript{b}, Karsten Kahl\textsuperscript{b} and Matthias Rottmann\textsuperscript{b}\\~\\
\small\textsuperscript{a}Department of Physics, University of Cyprus, PO Box 20537, 1678 Nicosia, Cyprus\\
\small\textsuperscript{b}Fakult\"at f\"ur Mathematik und Naturwissenschaften, Bergische Universit\"at Wuppertal\\
\small\textsuperscript{c}Computation-based Science and Technology Research Center, The Cyprus Institute\\~\\ 
       E-mail: \email{s.bacchio@hpc-leap.eu}}

\abstract{
We present  the Adaptive Aggregation-based Domain Decomposition Multigrid 
method extended to the twisted mass fermion discretization action.
We show comparisons of results as a function of tuning the parameters
that enter 
the twisted mass version of the DDalphaAMG library~\cite{Bacchio:DDalphaAMG}.
Moreover, we linked the DDalphaAMG library to the tmLQCD
software package and  give details on the performance of the multigrid solver
during HMC simulations at the physical point.
}

\FullConference{34th annual International Symposium on Lattice Field Theory\\
24-30 July 2016\\
University of Southampton, UK}

\begin{document}

\section{Introduction\label{sec:introduction}}

The Adaptive Aggregation-based Domain Decomposition Multigrid method, referred to as DD-$\alpha$AMG, 
has been introduced in Ref.~\cite{Frommer:2013fsa} as a solver for the Wilson clover operator, 
$D_W$. In DD-$\alpha$AMG a flexible iterative Krylov solver is preconditioned at every iteration step by a multigrid approach given by the error propagation
\begin{equation}\label{eq:error_propagation}
\epsilon\,\leftarrow\,\left(I-MD\right)^k\left(I-PD_c^{-1}P^\dagger D\right)\left(I-MD\right)^j\epsilon,
\end{equation}
where $M$ is the smoother, $j$ and $k$ are the number of pre- and post-smoothing iterations respectively, $P$ is the interpolation operator and $D_c = P^\dagger D_W P$ is the coarse grid operator.
The multigrid preconditioner exploits domain decomposition 
strategies having for instance as a smoother the Schwarz Alternating Procedure (SAP)~\cite{Luscher:2003qa}
and as a coarse grid correction an aggregation-based coarse grid operator. 
The method is designed to deal efficiently with both, infrared (IR)- and ultra-violet (UV)-modes of $D_W$. 
Indeed the smoother reduces the error components belonging to the UV-modes~\cite{Frommer:2013fsa}, while the
coarse grid correction deals with the IR-modes. This is achieved by using a interpolation operator $P$, which approximately 
spans the eigenspace of the small eigenvalues. Thanks to the property of local coherence~\cite{Luscher:2007se} 
the subspace can be approximated by aggregating over a small set of $N_v\simeq\mathcal{O}(20)$ test vectors $v_i$,
which are computed in DD-$\alpha$AMG via an adaptive setup phase~\cite{Frommer:2013fsa}.
We remark that the interpolation operator in DD-$\alpha$AMG
is $\Gamma_5$-compatible, i.e.~$\Gamma_5P = P\Gamma_{5,c}$.
Thanks to this property the $\Gamma_5$-hermiticity of $D_W$ is
preserved on the coarse grid as well -- i.e.~$D_c^\dagger = \Gamma_{5,c}D_c\Gamma_{5,c}$.

\begin{figure}
	\centering
	\resizebox{0.57\textwidth}{!}{\large \input{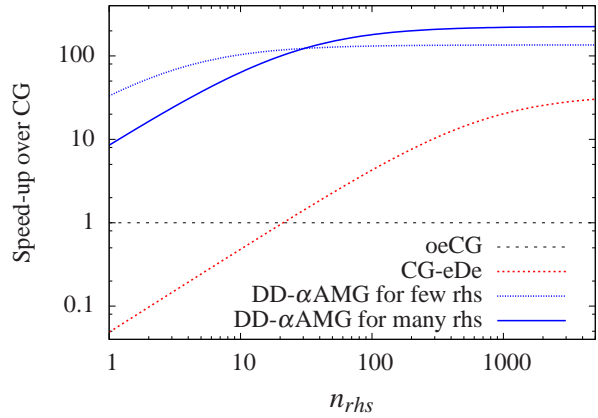}}
	\caption{
		\label{fig:comparisonCG} Speed-up of the DD-$\alpha$AMG 
		solver compared to odd-even preconditioned CG (oeCG) 
		and CG with exact deflation (CG-eDe) by using 1600 eigenvectors.
		The timings for the DD-$\alpha$AMG method and CG-eDe include 
		the time for the build-up and  the setup. The setup is required just once and then applied to  several
		rhs  on the same configuration. The optimized performance of DD-$\alpha$AMG
		for few and many rhs are obtained by changing the setup procedure, i.e.~three 
		setup iterations in case of few rhs and five setup iterations for many rhs,
		see also Ref.~\protect\cite{Alexandrou:2016izb}. 
	}
\end{figure}

Recently, we extended the DD-$\alpha$AMG approach to 
the Twisted Mass (TM) fermions discretization
with the TM operator $D(\pm\mu) = D_W \pm i \Gamma_5 \mu$~\cite{Alexandrou:2016izb}. 
When the PCAC (partial-conserved axial current) mass is tuned to zero,
the TM discretization cancels the linear lattice discretization effects \cite{Frezzotti:2003ni}.
The squared operator $D^\dagger(\mu) D(\mu) = D_W^\dagger D_W + \mu^2$ is
bounded from below by~$\mu^2$. Thus a finite twisted mass term $i \Gamma_5 \mu$ protects 
the TM operator $D(\mu)$ from being singular, unlike the Wilson
clover operator $D_W$ where this can happen for small quark masses.

By extending DD-$\alpha$AMG to the TM discretization
we observe a significant increase of the iteration count of the multigrid method 
at the physical value of the pion mass -- i.e.~at small values of $\mu$.
We find that the eigenvalue density of the squared operator is 
densely populated close to $\mu^2$~\cite{Alexandrou:2016izb}.
This increases significantly the iteration count on the coarse grid and slows down the method. 
By increasing the TM parameter on the coarsest grid
$\mu_{\rm coarse}=\delta\cdot\mu$ with $\delta \geq 1$
the coarse grid iteration count can be reduced by an order
of magnitude while simultaneously the fine grid iteration count only increases 
slightly. For $\delta \sim 5$ this improves the time to solution by a factor 4
for configurations of the $cA2.09.48$~\protect\cite{Abdel-Rehim:2015pwa} 
with lattice size $48^3\times96$ and pion mass $\sim 0.131$~GeV. We are thus
able to  achieve 
a similar speed up like it is found in the case of the Wilson clover operator at near physical pion masses.
In Ref.~\cite{Alexandrou:2016izb} we have presented a thorough analysis of the solver parameters
achieving speed-ups of more than a magnitude in time compared
to the Conjugate Gradient (CG) algorithm,
e.g.~a speed-up of $\mathcal{O}(100)$ when the solutions 
of at least ten right hand sides (rhs) are needed.
The result is depicted in Figure~\ref{fig:comparisonCG}.

\begin{figure}
	\centering
	\resizebox{0.97\textwidth}{!}{\Large \input{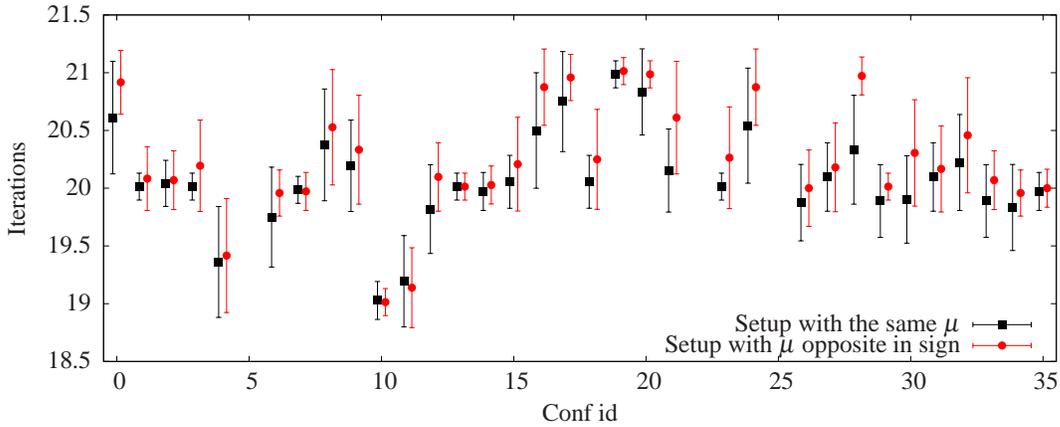}}
	\caption{\label{fig:observable}\label{fig:stability} Average iteration
		count on computing quark propagators for several configurations of $cA2.09.48$.
	}
\end{figure}

One advantage of the DD-$\alpha$AMG approach, when applied to the
TM operator, is the $\Gamma_5$-compatibility.
The TM term is still diagonal on the coarse grid similarly to the fine grid operator
and the coarse grid operator is given by $D_c (\mu) = D_c + i \mu \Gamma_{5,c}$.
Moreover, the same setup can be used for inverting both, $D(+\mu)$ and $D(-\mu)$, without affecting
the performance of the solver as shown in Fig.~\ref{fig:stability}.
Therefore, the method does not require an additional setup procedure
when linear systems with the squared operator $D^\dagger(\mu)D(\mu)$ have to be solved.

A code package containing the DD-$\alpha$AMG approach is publicly available 
in the DDalphaAMG library~\cite{Rottmann:DDalphaAMG}. 
Our TM version of the code is also publicly available in the resource~\cite{Bacchio:DDalphaAMG}
and we provide an interface to the tmLQCD software package~\cite{Jansen:2009xp} 
available at the resource~\cite{Finkenrath:tmLQCD}. In the next sections we give some 
details about the library and in section~\ref{sec:tmLQCD} we described in detail how the solver
can be called within tmLQCD.

\section{Overview of the DDalphaAMG library\label{sec:overview}}

The DDalphaAMG solver library has been released under GNU General Public
License at resource~\cite{Rottmann:DDalphaAMG}. This software package 
includes an implementation of the DD-$\alpha$AMG solver for the Wilson 
clover operator as described in Ref.~\cite{Frommer:2013fsa}. The
implementation is of production code quality, it includes
a hybrid MPI/\-openMP parallelization, state-of-the-art mixed precision
and odd-even preconditioning approaches and also SSE3 optimizations. 
Implementation details are available in Ref.~\cite{RottmannPhD}. 

Based on the DDalphaAMG code we have developed a version, which supports 
TM fermions, available at the resource~\cite{Bacchio:DDalphaAMG}. At the moment 
the following features have been added to the library: TM operator
with $N_f=2$ and twisted boundary conditions are supported, a different TM 
parameter can be applied on the even and odd sites -- required for the 
Hasenbusch mass preconditioning in the HMC simulations when odd-even 
preconditioning is used -- and a new interface to the library is provided.
All details about the interface can be found in the header of 
the library \texttt{DDalphaAMG.h} and a sample code, which links 
to the library is given in~\texttt{tests/DDalphaAMG\_sample.c}. 
Moreover the DDalphaAMG package can be used as an independent software package
including features as reading \texttt{LIME} configurations and reading/writing 
\texttt{LIME} vectors. More information can be found in 
the package documentation in the \texttt{doc/} directory.

The library interface is designed to provide an easy integration of the solver
to production codes, e.g.~codes which are used in computations of fermionic observables or
in HMC simulations. A minimal set of variables is required for the initialization 
of the solver, while a wide set of parameters is set by default
but can be modified for further optimizations. The list of parameters is given in the interface header, and information about their analysis and performance improvement is given in Refs.~\cite{Alexandrou:2016izb,Frommer:2013fsa}. 
The code conventions are the following: the $\gamma_5$ matrix is given by
\begin{equation}
 \gamma_5 = \begin{pmatrix}
 	$-$1 & \phantom{1} & \phantom{1} & \phantom{1} \\
 	  & $-$1           &             &  \\
 	  &             & 1        &  \\
 	  &             &             & 1\\
 \end{pmatrix}.
\end{equation}
while the representation of the other $\gamma_\nu$ matrices can be 
freely chosen, and sets are provided in \texttt{Clifford.h}. 
The order of the lexicographical index is fixed to $TZYX$.

The standard setup of the DD$\alpha$AMG approach is a three-level multigrid with aggregation block size of~$4^4$ 
between fine and first-coarse grid and of $2^4$ between first- and second-~(coarsest)
coarse grid. While the lower aggregation block size is automatically set,
the first aggregation block size can be further optimized by tuning at 
the same time the number of test vectors 
as it is described in Ref.~\cite{Alexandrou:2016izb}.

The number of levels and block sizes limit the maximal number 
of MPI parallel processes. At least two lattice sites 
of the coarsest grid are required per process when odd-even preconditioning is used.
Assuming the mentioned aggregation sizes 
the coarsest grid volume is given by $V/(4^4\cdot2^4)$ with $V$ the fine grid volume.
Thus the maximal number of MPI-processes is given by
$V/(2 \cdot4^4\cdot2^4)$. On recent machines, like Jureca,
we observe an almost ideal strong scaling up to the maximal number of processes. 
Moreover, it is possible to parallelize further by using openMP.
This increases the possible number of processes but without an ideal strong scaling. 
Additionally the library interface provides a trivial parallelization by splitting up the 
MPI-communicator. This can be used if solutions to several rhs have to computed in parallel 
when large number of processes are required.

\section{Employing the DD-$\alpha$AMG solver within tmLQCD\label{sec:tmLQCD}}
\label{sec:tmLQCD}

The DDalphaAMG library has been integrated into the tmLQCD software package~\cite{Jansen:2009xp},
which is commonly used by the European Twisted Mass (ETM) collaboration.
The code with the interface branch is available at the resource~\cite{Finkenrath:tmLQCD}. 
The DD-$\alpha$AMG solver can be used in all the applications of the software, which involve 
the inversion of the TM or the Wilson clover operator. 
The usage of the solver can be specified by \texttt{Solver = DDalphaAMG} in the tmLQCD-input-file
while the parameters can be set by adding a similar parameter environment like it is shown in the
Listing~\ref{lst:input_file}.

{\footnotesize
	\definecolor{grey}{rgb}{0.9,0.9,0.9}
\begin{lstlisting}[backgroundcolor = \color{grey}, multicols = 2, columns=flexible, keepspaces=true, captionpos = b, label = lst:input_file, caption = DDalphaAMG parameters in a tmLQCD input file]
 BeginDDalphaAMG
   MGBlockX = 4
   MGBlockY = 4
   MGBlockZ = 4
   MGBlockT = 4
   MGSetupIter = 5
   MGCoarseSetupIter = 3
   MGNumberOfVectors = 24
   MGNumberOfLevels = 3
   MGCoarseMuFactor = 5
 EndDDalphaAMG
 
 BeginOperator TMWILSON
   2kappaMu = 0.05
   kappa = 0.177
   Solver = DDalphaAMG
   SolverPrecision = 1e-14
 EndOperator
 
\end{lstlisting}}

A detailed description of the available parameters can be found in the package documentation in
\texttt{doc/main.pdf}. The solver is tested within the 
applications \texttt{invert} and \texttt{hmc\_tm}, while in any case an additional check
of the residual is performed in tmLQCD. 
Note that even if the solution of the odd-even preconditioned operator is required, 
the DD-$\alpha$AMG approach uses the full operator 
for the inversion. Since odd-even preconditioning reduces the sparsity of the operator, in DD-$\alpha$AMG it is used only on the coarsest level and in the smoother.
This is also the case if an inversion of the squared operator is required,
where DDalphaAMG performs two inversions of the non-squared operator. 

\subsection{HMC simulation with DDalphaAMG}

\begin{figure}
	\centering
	\resizebox{0.57\textwidth}{!}{\large \input{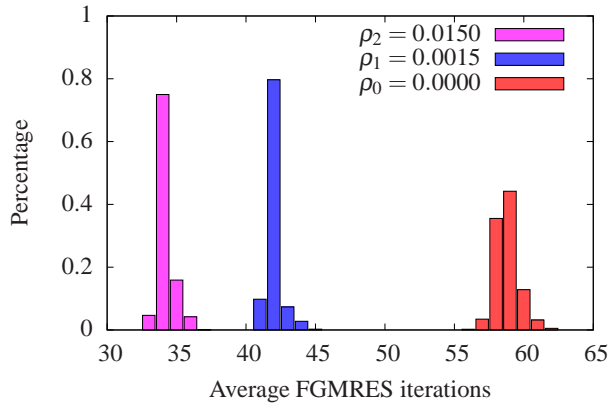}}
	\caption{\label{fig:hmc} FGMRES iteration counts averaged over the trajectory when the solver 
		is used for computing the force terms in the HMC simulation. The data are for the ensemble $cA2.09.64$ with 
		statistics of 2000 trajectories. The parameters $\rho_i$ set the mass for the Hasenbusch preconditioning. 
		The iteration counts include two inversions of the non-squared TM operator.
	}
\end{figure}

We perform an HMC simulation to generate a $64^3\times128$ ensemble at physical pion mass with an integration scheme equivalent 
to the one used in Ref.~\cite{Abdel-Rehim:2015pwa} for the ensemble $cA2z.09.48$. 
The DD-$\alpha$AMG solver is employed in the force term computation, heat-bath and acceptance step.
In the integration scheme the Hasenbusch mass preconditioning~\cite{Hasenbusch:2002ai} is used.
During the integration the squared operator $\hat D^\dagger(\mu) \hat D(\mu) + \rho_i^2$ has to be inverted, 
where $\hat D(\mu)$ is the odd-even reduced or odd-even preconditioned TM operator. 
In DDalphaAMG this is done by first inverting $\hat D(\mu) + i \rho_i \hat\Gamma_5$
and then $\hat\Gamma_5\hat D(-\mu)\hat\Gamma_5 - i \rho_i \hat\Gamma_5$, 
where $\hat\Gamma_5$ is $\Gamma_5$ restricted to the odd lattice sites. 
The same interpolation operator $P$ can be used for all inversions of
the operators involved in the HMC procedure. 
The setup is built once at the beginning of each trajectory, 
where three iterations on the fine grid and three on the first coarse grid are used.
During the integration the setup is updated every time before the operator
with the smallest Hasenbusch mass $\rho_0=0$ is inverted. The update
is done by one fine grid and three first coarse grid iterations.
By using this approach the DD-$\alpha$AMG solver showed very stable iteration counts for all $\rho_i$
as it is depicted in Figure~\ref{fig:hmc}.

The calculation of the force term requires a higher accuracy for the inversions when the DD-$\alpha$AMG approach 
is used instead of the CG solver. This is required to maintain the reversibility of the HMC.
Indeed, while the solution provided by CG only depends
on the current configuration and the right hand side, the multigrid
setup carries information from the previous configurations which thus
influences the solution corresponding to the current configuration.

\section{Conclusions and outlook\label{sec:conclusions}}

The DD-$\alpha$AMG approach is extended to the case of $N_f=2$ twisted mass fermions~\cite{Alexandrou:2016izb}. 
The code is publicly available in the twisted mass version of the DDalphaAMG library~\cite{Bacchio:DDalphaAMG}.
Moreover, we implement an interface in tmLQCD available under~\cite{Finkenrath:tmLQCD}.
After tuning the parameters, the inversions are performed more than two orders of magnitude faster as 
compared to standard CG. Within the HMC simulations with $N_f=2$, DD-$\alpha$AMG achieves a speed-up of an 
order of magnitude compared to standard CG. Future steps will be the integration of the heavy quark twisted mass 
operator into the DDalphaAMG library~\cite{Bacchio:DDalphaAMG}. Furthermore, we plan to update the vectorization to
AVX instructions.

\subsection*{Acknowledgments}
This project has received funding from the  Horizon 2020 research and innovation program of the European Commission
under the Marie Sklodowska-Curie grant agreement No 642069. S.B.~is supported by this program.
We would like to thank Giannis Koutsou for fruitful discussions and for helping us accessing the configurations.
We also thank Artur Strebel and Simon Heybrock for guidance with the DD-$\alpha$AMG code and Bj\"orn
Leder for his suggestion to shift the twisted mass on the coarse grid
to speed-up the coarse grid solver. The authors gratefully acknowledge 
the Gauss Centre for Supercomputing e.V.~for funding the project \emph{pr74yo} by 
providing computing time on the GCS Supercomputer SuperMUC at Leibniz Supercomputing Centre,
the computing time granted by the John von Neumann Institute for Computing 
(NIC) and provided on the supercomputer JURECA at J\"ulich Supercomputing Centre (JSC) through the grant \emph{ecy00}, 
the High Performance Computing Center in Stuttgart for providing computation time on the High
Performance Computing system Hazel Hen through the grant \textit{GCS-Nops
(44066)}, and the computational resources on Cy-Tera at the Cyprus Supercomputing Center through the grant \emph{lspre258s1}.

\end{document}